# Nutritional blood concentration biomarkers in the Hispanic Community Health Study/Study of Latinos: Measurement characteristics and power

Lillian A. Boe[1,*], Yasmin Mossavar-Rahmani[2], Daniela Sotres-Alvarez[3], Martha L. Daviglus[4], Ramon A. Durazo-Arvizu[5], Bharat Thyagarajan[6], Robert C. Kaplan[2,7], Pamela A. Shaw[8]

[1] Department of Biostatistics, Epidemiology, and Informatics, University of Pennsylvania Perelman School of Medicine, Philadelphia, Pennsylvania [2] Department of Epidemiology and Population Health, Albert Einstein College of Medicine, Bronx, New York [3] Department of Biostatistics, Gillings School of Global Public Health, University of North Carolina at Chapel Hill, Chapel Hill, North Carolina [4] Institute for Minority Health Research, University of Illinois at Chicago, Chicago, Illinois [5] Department of Public Health Sciences, Loyola University Chicago Health Sciences Campus, Chicago, Illinois [6] Department of Laboratory Medicine and Pathology, University of Minnesota, Minneapolis, Minnesota [7] Public Health Sciences Division, Fred Hutchinson Cancer Research Center, Seattle, Washington [8] Biostatistics Unit, Kaiser Permanente Washington Health Research Institute, Seattle, Washington

## Abstract

Measurement error is a major issue in self-reported diet that can distort diet-disease relationships. Use of blood concentration biomarkers has the potential to mitigate the subjective bias inherent in self-report. As part of the Hispanic Community Health Study/Study of Latinos (HCHS/SOL) baseline visit (2008-2011), self-reported diet was collected on all participants (N=16,415). Blood concentration biomarkers for carotenoids, tocopherols, retinol, vitamin B12 and folate were collected on a subset (N=476), as part of the Study of Latinos: Nutrition and Physical Activity Assessment Study (SOLNAS). We examine the relationship between biomarker levels, self-reported intake, Hispanic/Latino background, and other participant characteristics in this diverse cohort. We build regression calibration-based prediction equations for ten nutritional biomarkers and use a simulation to study the power of detecting a diet-disease association in a multivariable Cox model using a predicted concentration level. Good power was observed for some nutrients with high prediction model $R^2$ values, but further research is needed to understand how best to realize the potential of these dietary biomarkers. This study provides a comprehensive examination of several nutritional biomarkers within the HCHS/SOL, characterizing their associations with subject characteristics and the influence of the measurement characteristics on the power to detect associations with health outcomes.

Introduction

Measurement error is a major issue in self-reported diet.[1] Specifically, systematic and random error in self-reported diet can bias exposure-disease associations, with the former causing either attenuation or inflation bias.[2] There has been growing interest in using nutritional biomarkers to better capture aspects of diet associated with health outcomes.[3-6] Two major classes of dietary biomarkers are recovery and concentration biomarkers.[7-8] Recovery biomarkers are quantitatively related to intake and have been established as objective (unbiased) biomarkers for intake, such as doubly labeled water for energy and urinary nitrogen for protein,.[7] Blood concentration biomarkers are more readily available for a wider range of nutrients, for example, folate, tocopherols, carotenoids and vitamin B12.[8] Serum and plasma concentrations do not measure absolute levels of intake and typically have lower correlations with food or nutrient intakes compared to those expected for recovery biomarkers.[9] Concentration markers have been advocated as useful for establishing diet-disease relationships, as a practical approach to obtain measures that may capture aspects of diet distinct from those measured by self-report.[3,5] Additional work is needed to explore how concentration levels relate to disease incidence in Hispanic/Latino groups. Throughout this manuscript, we will refer to serum and plasma concentrations as blood *concentration biomarkers*.

Previous studies reported that cultural differences in Hispanics/Latinos in the United States (US), including Hispanic country of origin (background), language preference, acculturation, and education, contribute to differences in measurement characteristics of micronutrients from 24-hour recalls.[10-13] The measurement error in micronutrient intake from self-reported dietary intake instruments has not been adequately investigated in Hispanic/Latino groups in the US. In the Study of Latinos: Nutrition and Physical Activity Assessment Study (SOLNAS), blood concentration biomarkers and participant characteristics were collected in a subsample of 476 participants from the Hispanic Community Health Study/Study of Latinos (HCHS/SOL) (N = 16,415). We describe how these blood concentration biomarkers vary with participant characteristics, such as sex, age, and background. We also examine the relationship between these concentration biomarkers and their corresponding self-reported, 24-hour dietary recall measures, which have not yet been investigated in this population. We build prediction models for each concentration marker, which elucidate the strength of association between concentration levels and 24-hour recalls. We develop prediction equations to implement regression calibration, a method that replaces an unobserved true exposure (e.g., average blood concentration level) by an estimate based on other measured covariates[14, 15]. Studies of concentration biomarker levels and their measurement error properties do not yet exist in this population of Hispanics/Latinos. This work is important as calibration studies developed in other contexts are not expected to be transportable to this population.[16] Combining data from blood concentration biomarkers with self-reported supplement and dietary intake data is a potential strategy to obtain predictions of usual concentration level, which in turn may be used as an alternative estimate of diet-disease associations that is not subject to errors from reporting bias, problematic questionnaires, or insufficient nutrient databases.[3]

We illustrate how our prediction models could be used to impute missing concentration levels in HCHS/SOL participants not part of SOLNAS . These imputed values allow for the concentration levels to be included in outcome-diet association studies in the larger cohort, similar to analyses that have been done using recovery biomarkers.[17,18] Such an analysis could be used to assess whether the usual concentration level is associated with a disease outcome. We use numerical simulation to study the statistical power of detecting associations between true average serum or plasma measures and a hypothetical time-to-event outcome using imputed concentration levels. Our simulation study offers a new approach for evaluating the potential of concentration biomarkers measured only on a subset as the exposure in a health outcome model.

Subjects and methods

**Study population**

HCHS/SOL is a community-based cohort study of 16,415 men and women aged 18-74 years who self-identified as Hispanic/Latino and were recruited in 2008-2011 from randomly selected households in four US metropolitan areas (Bronx, NY; Chicago, IL; Miami, FL; San Diego, CA). The selected sites constitute 4 of the 11 urban centers in the US with the largest number of Hispanics/Latinos.[19] The primary objective of HCHS/SOL is to study associations between baseline risk factors and chronic conditions in a cohort of diverse Hispanics/Latinos.[8] Study participants underwent a comprehensive baseline clinical examination that included biological, behavioral, and social-demographic assessments, yearly telephone follow-up, a second (2014-2017) and third clinic visit (2020-2023). Individuals were invited to participate in the SOLNAS sub-study within 7 to 12 months of their initial HCHS/SOL clinic visit. Participants were excluded for having any medical condition that might affect weight stability or biomarker performance or having extended travel planned during the study period.[1] A total of 1,360 HCHS/SOL participants were invited to enroll in SOLNAS; 485 (35.7%) consented to participate in 2010-2012. Of these 485 enrolled individuals, 7 dropped out and 2 did not provide blood biomarker data, leaving 476 participants and a subsample of 98 that repeated the study protocol after 6 months to provide reliability information (see Web Figure 1). We briefly outline the study protocol here; Mossavar-Rahmani et al. (2015) provides further details.[1]

**Study protocol and procedures**

The SOLNAS study protocol consisted of two clinic visits, with in-home activities between visits. The doubly labeled water (DLW) and urinary nitrogen recovery biomarkers were used to assess total energy expenditure and protein intake, respectively[1]. At Visit 1, participants arrived following a 4-hour fast and provided pre- and post-DLW spot urine samples and completed a 24-hour dietary recall. At Visit 2, twelve days later, participants completed the DLW protocol and provided a fasting blood draw and urine specimen. Ninety-eight participants (20%) repeated the study for visits 3 and 4. Plasma and serum samples were stored at -70 degrees until analysis.

*Serum & Plasma Biomarkers*

Assays were completed by the HCHS/SOL central laboratory at University of Minnesota Fairview Hospital Laboratory for total cholesterol, HDL, LDL, blood concentrations of α-carotene, β-carotene, α-tocopherol, γ-tocopherol, vitamin B12, β-cryptoxanthin, retinyl palmitate, folate, lycopene, retinol, and zeaxanthin. Prior studies suggested that this subset of micronutrients were useful biomarkers associated with fruit and vegetable intake and may have a preventive role in chronic disease.[1,7, 20-21] Blinded duplicates were analyzed for a 10% quality control subsample; Web Table 1 presents intraclass correlation coefficients (ICC) and coefficient of variation (CV). Agreement was high (CV <15%, ICC ~90% or higher) for all markers, except retinol and retinyl palmitate, the latter of which was often at or below the limit of quantification. Carotenoids and tocopherol were measured by HPLC using a slight modification of the methods[22,23], which allowed for simultaneous detection of both carotenoids and tocopherol. Vitamin B12 was measured in serum using the Siemens ADVIA Centaur Vitamin B12 assay (Siemens Healthcare Diagnostics, Deerfield IL). Folate was measured using the Siemens ADVIA Centaur Folate assay (Siemens Healthcare Diagnostics, Deerfield IL). Retinol was measured by a modification of the HPLC method[22] and calibration[23]. The assay modification includes the addition of 0.015% N, N-diisopropylethylamine in the HPLC solvent as an aid to analyte recovery. Serum cholesterol measures were analyzed by the same laboratory from a fasting blood draw taken at HCHS/SOL study baseline. Web Appendix 1 contains additional details.

*Dietary and participant characteristics assessment*

Two 24-hour dietary recalls were collected during the HCHS/SOL baseline period using the Nutrition Data System for Research (NDS-R) software (version 11); the first recall was collected in person at the clinic and the second by telephone within 3 months. In SOLNAS, an in-person 24-hour dietary recall was collected at Visit 1. Further details on these procedures have been described previously[1, 20, 24]. In Web Appendix 1, we include a few key details.

At the HCHS/SOL baseline visit, participants provided demographic, health, lifestyle, and acculturation characteristics, including self-reported physical activity using a modified Global Physical Activity Questionnaire (GPAQ) and self-reported 24-hour supplement intake. In Web Appendix 1, we define the supplement use, hypertension, high cholesterol, and diabetes variables.

<u>Statistical analyses</u>

To address skewness, we analyzed log-transformed biomarker measurements from SOLNAS, as is typical for these nutrients.[4, 7, 25] A 2-day mean obtained from the baseline telephone recall and the SOLNAS in-person recall was used to estimate usual nutrient intake. The mean of multi-day 24-hour recalls is commonly used to estimate usual diet and reduce within-person variation.[25, 26] We excluded two participants who reported energy intake < 500 kcal/day at their SOLNAS 24-hour recall. The data had a missing rate of <1% for most biomarker and 24-hour recall values (Tables 2 and 3). Geometric means and 95% CI (2.5$^{th}$-97.5$^{th}$ percentiles) were reported for each nutrient by sex and background.

Linear regression models predicting the underlying average concentration level of each nutrient were constructed by regressing the log-transformed biomarker measure on the 2-day mean of self-reported intakes and age, sex, background (Central American, Cuban, Dominican, Mexican, Puerto Rican, South American), language preference (Spanish, English), body mass index (BMI), education (less than high school, high school graduate/GED, some post high school, college graduate or higher), income (<$20,000, $20,001-$50,000, $50,000+, not reported), smoking status (current smoker or not), alcohol use (<1 drink/week, 1-7 drink/week, 7+ drink/week), diabetes, family history of diabetes, hypertension, high cholesterol, log total energy intake, supplement use (yes, no), and the associated self-reported intake and supplement use variables. Prediction models were built for all nutrients except retinyl palmitate, excluded due to the high percentage of biomarker missingness (46.4%). Each prediction model was adjusted by other nutrient-specific factors based on a priori knowledge (Web Table 2) [7-8, 27-31]. Details on the specific variables of interest for each nutrient, including cholesterol-related variables for carotenoids and tocopherols, are provided in Web Appendix 2. Web Table 3 presents a mapping between biomarkers and self-reported nutrients for each prediction model. Most biomarkers had an exact self-reported nutrient match, except zeaxanthin, for which the closest corresponding self-reported measure was lutein plus zeaxanthin. We use a significance level of 0.05 to quantify strong associations in the multivariable linear model.

Since our statistical analysis approach was to build prediction models for the underlying average concentration level, we do not perform variable selection and consider the full model our gold standard. We check for overfitting and investigate variable redundancy using two approaches proposed by Harrell (2015)[14]; see details in Web Appendix 3. We report $R^2$ values; Prentice $R^2$ values[32], i.e. $R^2$ values that account for the within-person variability in the biomarker; partial $R^2$ values; correlations between biomarkers and their corresponding self-reported measures; and new $R^2$ measures that could exist if there were 2 and 4 repeat measures of each biomarker available on all SOLNAS participants (Web Table 5). The latter measures show the potential to increase the precision of the biomarker by reducing variance from day-to-day diet fluctuations.

We conduct simulations based on a few key demographic and dietary variables in the HCHS/SOL cohort to study the power of detecting the association between true average concentrations and time to incident diabetes. To represent the varying predictive accuracy observed in the SOLNAS data, we simulate $\beta$-Cryptoxanthin, lycopene, and folate which had high, medium, and low $R^2$ values, respectively. For simplicity, we assume two additional covariates, age and BMI, are also related to the health outcome of interest. We assumed a Cox Proportional Hazards model and chose a hazard ratio (HR) for the true average exposure that would have approximately 90% power for the HCHS/SOL simulation settings. While numerical formulas are available for power calculations in the Cox model, they are less straightforward for multivariable regression scenarios with error-prone exposures. Web Appendix 4 contains details of the numerical simulation, including the variable distributions, measurement error model, and outcome model.

We present mean and median percent biases, average standard errors (ASE), empirical standard errors (ESE), 95% coverage probabilities (CP), and power across 1000 simulations for the estimated HR for simulated $\beta$-Cryptoxanthin, lycopene, and folate. We compare power for six analysis approaches: (1) true average concentration level, (2) naïve biomarker, (3) calibrated biomarker, (4) naïve self-report, (5) calibrated self-report, and (6) optimal combination of the calibrated biomarker and calibrated self-report. The optimal combination approach is based on a method proposed by Spiegelman et al. (2001)[33], which computes a generalized inverse-variance weighted average of the calibrated self-report and biomarker. We use the regression calibration approach introduced by Prentice (1982)[15].

Results

Baseline demographic and lifestyle characteristics of the 476 participants who had data on at least one blood biomarker and the 16,415 HCHS/SOL study participants are shown in Table 1. Baseline characteristics of SOLNAS participants are similar to those of HCHS/SOL. In both HCHS/SOL and SOLNAS, 60% of participants were female, 40% had a BMI > 30, and the mean age at baseline was 46.0 years. SOLNAS participants were 29.8% Mexican, 25.8% Puerto Rican, 14.5% Cuban, 10.7% Central American, 10.1% Dominican, and 9.0% South American. Spanish language preference was 76.5% for SOLNAS and 79.9% for HCHS/SOL, with approximately 75% of both groups making under $50,000 and having less than a college education.

Table 2 presents age-adjusted geometric means for nutritional biomarkers and 24-hour recall measures by sex for the mean SOLNAS age of 46.1 years. In Table 3, we show age-and-sex-adjusted geometric means for all biomarkers and 24-hour recall measures by Hispanic/Latino background. The results suggest that self-report and biomarker measures differed by both sex and Hispanic/Latino background. Table 2 shows that blood biomarker levels of α-carotene and β-cryptoxanthin were quite different for females compared to males, a trend not present in the self-reported values. For other nutrients, like retinol and zeaxanthin, small differences were observed between females and males for both the biomarker and self-report. In Table 3, we observe that some Hispanic/Latino backgrounds with high mean levels of blood concentrations also had high mean self-reported intakes. This was true for β-carotene, where South Americans had the highest geometric mean of self-report intakes and correspondingly had large geometric means of biomarker values. For other nutrients, such as lycopene, we did not see this same alignment. Participants of Mexican heritage, the group with the highest mean intake of lycopene based on the 24-hour recall, did not exhibit high biomarker levels compared to participants of other backgrounds.

The fitted regression calibration (prediction) model coefficients for the logarithm of biomarker measures are shown in Table 4 for selected variables of interest. For all nutrients, Hispanic/Latino background, supplement use, age, and body mass index were important independent calibration model predictors. For a subset of nutrients including α-carotene, β-carotene, β-cryptoxanthin, folate, and zeaxanthin, the corresponding self-reported measure was a useful predictor of the concentration level, as evidenced by strong associations in the multivariable model (p-value<0.03). For the remaining nutrients, the self-report had little predictive value. We found no conclusive evidence of overfitting or variable redundancy (data not shown). As a sensitivity analysis, Web Table 4 presents variables selected for the reduced model by stepwise selection using Akaike Information Criterion (AIC). Reduced models for α-carotene, β-carotene, β-cryptoxanthin, folate, zeaxanthin, and lycopene kept the self-report as a predictor. This is consistent with our observations from the full models, where we saw strong associations between the self-report and biomarker for the same nutrients except lycopene.

In Web Table 5, we present $R^2$ values based on the full prediction models. The $R^2$ values in decreasing order are: β-cryptoxanthin, 0.5035; α-carotene, 0.4692; β-carotene, 0.4562; α-tocopherol, 0.3486; zeaxanthin, 0.2728; γ-tocopherol, 0.2340; lycopene, 0.2196; folate, 0.1717; retinol, 0.1185; and B12, 0.1013. Partial $R^2$ values for the self-report were highest for α-carotene, 0.0725; β-cryptoxanthin, 0.0699; zeaxanthin, 0.0373; and β-carotene, 0.0326; with values ≤ 0.0137 for all other nutrients. Similarly, 24-hour recalls of β-cryptoxanthin, α-carotene, β-carotene, and zeaxanthin had the highest correlations with their pertinent concentration markers, ranging from 0.24 to 0.41. For α-tocopherol, γ-tocopherol, lycopene, folate, retinol, and B12, these correlations ranged from 0.02 to 0.12, suggesting weak-to-little relationship. For nutrients like α-tocopherol, we notice we can appreciably improve the $R^2$ by having more than one biomarker measure, as the theoretical $R^2$ values for two and four repeat biomarker measures are 0.4219 and 0.4700, respectively, getting closer to the ideal Prentice $R^2$ of 0.5304.

Figures 1 and 2 show comparisons between the main and reliability study measures for the biomarker and the self-reported intakes, respectively, at visits 1 and 3. Each plot also presents the ICC, or the Pearson correlations between the repeat measures. For the biomarkers, correlations were high (> 0.6) for β-cryptoxanthin (0.89), β-carotene (0.87), $\alpha$-carotene (0.81), B12 (0.79), zeaxanthin (0.70), lycopene (0.68), $\alpha$-tocopherol (0.66), and $\gamma$-tocopherol (0.62); moderate (0.3 to 0.6) for folate (0.59), and low (< 0.3) for retinyl palmitate (0.17) and retinol (0.09). The correlations for the self-reported 24-hour recall measurements were generally not as high, possibly due to the time between recalls: moderate for folate (0.46) and low for α-tocopherol (0.29), γ-tocopherol (0.27), lutein plus zeaxanthin (0.27), B12 (0.26), retinol (0.23), β-carotene (0.22), α-carotene (0.18), β-cryptoxanthin (0.14), and lycopene (0.07).

Table 5 presents results from our simulation study. Models using the naïve self-report are estimated to have mean percent biases of -88.82% for $\beta$-cryptoxanthin, -98.95% for lycopene, and -91.047% for folate. We also see appreciable bias in the naïve biomarker for each nutrient, though not as extreme. The calibrated biomarker, calibrated self-report, and optimal combination reduce the mean percent biases to under 5% in all scenarios. For $\beta$-cryptoxanthin, 95% coverage probability is maintained for both calibration approaches and the optimal combination. Power improves slightly from the naïve self-report (0.689) or calibrated self-report (0.679) using the optimal combination approach (0.698). Note due to computational limitations, power and CP for the optimal approach are calculated from a Wald confidence interval computed using bootstrap standard errors, which can be problematic due to the potential asymmetry of the bootstrap distribution[34]. This likely explains the results observed for simulations based on lycopene and folate, where the optimal approach resulted in slightly lower power compared to the naïve self-report. Further, CP values were greater than 95% for the optimal combination in simulations of lycopene (CP = 0.963) and folate (CP = 0.969).

Discussion

Self-reported diet has been linked to chronic diseases, however, nutritional biomarkers of intake confirming these associations are lacking. Many epidemiologic studies rely on self-reported dietary intakes as their primary exposure of interest, which is problematic due to the generally substantial measurement error in subjective, self-reported dietary assessment. It is therefore of interest to consider biological indicators of nutritional intake, such as concentration biomarkers, which may capture an aspect of dietary exposure important for health outcomes and avoid the subjective biases known to influence self-reported diet. Our study is the first to describe the measurement error structure of serum and plasma concentration biomarkers and corresponding self-reported 24-hour dietary recall measures in a diverse cohort of US Hispanic/Latino adults. Our contribution is twofold; we first examined the levels of blood biomarkers in this cohort and we also assess how useful these biomarkers may be in evaluating diet-disease associations in the HCHS/SOL.

Previous studies that investigate dietary components using HCHS/SOL have been limited to nutrients such as energy, protein, sodium and potassium, for which an associated recovery biomarker is available.[1,20] We found that levels of blood concentration biomarkers varied by participant characteristics. Specifically, levels of some carotenoids were higher for women compared to men, a difference that was not observed in the corresponding self-reported intakes.

Similarly, we observed low serum levels of lycopene in participants of Mexican heritage compared to participants of other backgrounds, despite that this group self-reported the highest mean intake of lycopene. Some genetic factors affecting metabolic differences in biomarker levels by background have been noted previously[35]. Serum concentration biomarkers and self-reported intakes are likely capturing different quantities, both having components of bias that limit their ability to assess dietary intake. We examine the relationship between blood biomarkers and self-reported diet and found self-report was a strong predictor of concentration level for α-carotene, β-carotene, β-cryptoxanthin, and zeaxanthin. For these nutrients, correlations between the self-report and corresponding biomarker were ≥ 0.24 and partial $R^2$ values for the self-report were highest, ranging from 0.0326 to 0.0725. Overall, concentration biomarkers except retinol had high ICC measures, ranging from 0.59 to 0.89, likely due to a combination of an accurately recorded assay and low biological variability. Self-reported ICC measures were much lower, ranging from 0.07 to 0.46, which is expected due to the high variability in within-person 24-hour recall measures of diet, particularly those measured further apart in time[25].

Our analysis approach of building prediction equations from concentration biomarkers differed from previously published results in other cohorts. Lampe et al. (2017)[7] built prediction equations for micronutrients based on a feeding study from the Women's Health Initiative (WHI) by regressing dietary intake on concentration biomarkers. Prentice et al. (2020)[6] augmented four of these prediction equations by adding self-reported food frequency questionnaire (FFQ) intake as a covariate to assess the change in $R^2$ values. In contrast, we regressed the concentration level on self-reported dietary intake and other covariates. Studies from the WHI reported higher $R^2$ values for some nutrients, like $\alpha$-carotene ($R^2$=0.47 for SOLNAS vs. $R^2$=0.53 for Lampe et al. 2016) and folate (0.17 vs. 0.49). By design, feeding study participants may have lower day-to-day variability than HCHS/SOL participants who are participating in an observational study and reporting usual diet, contributing to lower $R^2$ measures for SOLNAS.

Prentice et al. (2019)[4] observed that in the WHI feeding study context[7], concentration biomarkers with an $R^2 \geq 0.36$ were just as correlated with their corresponding true intake values as recovery biomarkers, and thus proposed this cutoff for identifying useful concentration markers. $R^2$ values ranged from 0.27 to 0.50, which are within the range of $R^2$ values observed for well-accepted recovery biomarkers in SOLNAS (Mossavar-Rahmani et al. 2015)[1]. One important criterion for assessing concentration biomarkers is whether using the predicted usual concentration level in the outcome model allows for good power to detect a diet-disease association. For our exposure variable simulated to represent $\beta$-cryptoxanthin ($R^2$ = 0.5035), there was reasonable power (~70%) to detect a 20% increase in blood concentration level. The simulations of lycopene and folate indicated that as prediction model $R^2$ values decrease, power fell short (< 18%). Our paper offers a new approach for assessing the usefulness of concentration biomarkers in the biomarker sub-study context.

Our study had some limitations. Some variables in the prediction equations, namely the cholesterol variables, one dietary recall for the 2-day mean, and supplement intake, come from the HCHS/SOL baseline visit rather than SOLNAS. Self-reported diet coincident with the blood biomarkers may lead to stronger correlations. A limitation of our simulation study is that CP and power for the optimal combination approach are based on bootstrap percentile or Wald confidence intervals, which can be problematic in some finite-sample settings[35]. Due to complexity of two-stage regression in a multivariable semiparametric Cox model, there aren't simple analytical formulas available to inform sample size and power calculations. Further work is needed in this area. Analyses were exploratory and did not adjust for multiple comparisons.

An important finding of this study is that levels of blood concentration biomarkers often varied by participant characteristic and showed different intake patterns compared to their corresponding self-reported measures. We showed that moderate correlations between biomarkers and 24-hour recall measures resulted in prediction models where the self-report was a useful predictor of usual concentration level. Our study also revealed that using a predicted concentration level can be useful for estimating diet-disease associations in cohorts of similar size to HCHS/SOL, particularly when the prediction model $R^2$ is reasonably high. While the low $R^2$ of other micronutrients led to reduced power in the association studies, these nutrients have potential and we hope that our paper stimulates interest in future studies that aim to obtain repeat concentration biomarkers on substudy participants. Repeat biomarkers might increase the prediction model $R^2$ and improve the power of detecting outcome associations with average concentration markers. The authors expect that the current investigation of serum and plasma biomarkers will inform future cohort studies that explore diet-disease associations when covariate measurement error is present.

Acknowledgements

This work was supported in part by NIH grant R01-AI131771. The Study of Latinos: Nutrition & Physical Activity Assessment Study (SOLNAS) was supported by grant R01HL095856 from the National Heart, Lung, and Blood Institute. The Hispanic Community Health Study/Study of Latinos was carried out as a collaborative study supported by contracts from the National Heart, Lung, and Blood Institute to the University of North Carolina (NO1-HC65233), University of Miami (N01-HC65234), Albert Einstein College of Medicine (N01-HC65235), Northwestern University (N01-HC65236), and San Diego State University (N01-HC65237). The following contribute to the Hispanic Community Health Study/ Study of Latinos through a transfer of funds to the National Heart, Lung, and Blood Institute: the National Center on Minority Health and Health Disparities, the National Institute of Deafness and Other Communications Disorders, the National Institute of Dental and Craniofacial Research, the National Institute of Diabetes and Digestive and Kidney Diseases, the National Institute of Neurological Disorders and Stroke, and the Office of Dietary Supplements. Additional support at the Albert Einstein College of Medicine was provided from the Clinical and Translational Science Award (UL1 TR001073) from the National Center for Advancing Translational Sciences at the National Institutes of Health.

The authors would like to thank the investigators of the HCHS/SOL study for the use of their data. A list of HCHS/SOL investigators, managers and coordinators by field center can be found here: https://sites.cscc.unc.edu/hchs/Acknowledgement.

Data Availability Statement

The data used in this paper were obtained through submission and approval of a manuscript proposal to the Hispanic Community Health Study/Study of Latinos Publications Committee, as described on the HCHS/SOL website. For more details, see https://sites.cscc.unc.edu/hchs/publications-pub.

**Table 1.** Demographic and other personal characteristics in the Study of Latinos: Nutrition & Physical Activity Assessment Study (SOLNAS) (N=476) and Hispanic Community Health Study/Study of Latinos (HCHS/SOL) (N=16415).

| Characteristic | SOLNAS[2] N (%) | HCHS/SOL[3] N (%) |
|---|---|---|
| Age | | |
|     18-44 years | 182 (38.2) | 6701 (40.8) |
|     45+ years | 294 (61.8) | 9714 (59.2) |
| Sex | | |
|     Male | 187 (39.3) | 6580 (40.1) |
|     Female | 289 (60.7) | 9835 (59.9) |
| Background[1] | | |
|     Central American | 51 (10.7) | 1732 (10.6) |
|     Cuban | 69 (14.5) | 2348 (14.3) |
|     Dominican | 48 (10.1) | 1473 (9.0) |
|     Mexican | 142 (29.8) | 6472 (39.4) |
|     Puerto Rican | 123 (25.8) | 2728 (16.6) |
|     South American | 43 (9.0) | 1072 (6.5) |
|     Other/More than one | 0.0 (0.0) | 503 (3.1) |
| Language Preference | | |
|     English | 112 (23.5) | 3296 (20.1) |
|     Spanish | 364 (76.5) | 13119 (79.9) |
| Body Mass Index (kg/m$^2$) | | |
|     Underwt/normal (<25) | 96 (20.2) | 3321 (20.2) |
|     Overweight (25-30) | 186 (39.1) | 6116 (37.3) |
|     Moderately obese (30-35) | 118 (24.8) | 4219 (25.7) |
|     Morbidly Obese (35+) | 76 (16.0) | 2688 (16.4) |
| Education | | |
|     Less than High School | 153 (32.1) | 6207 (37.8) |
|     High School Graduate/GED[1] | 119 (24.9) | 4180 (25.5) |
|     Some Post High School | 70 (14.7) | 2053 (12.5) |
|     College graduate or higher | 135 (28.3) | 3884 (23.7) |
| Income | | |
|     < $20,000 | 231 (48.5) | 7207 (43.9) |
|     $20,001-$50,000 | 137 (28.8) | 5058 (30.8) |
|     $50,000+ | 66 (13.9) | 2662 (16.2) |
|     Not reported | 42 (8.8) | 1488 (9.1) |
| Currently employed[1] | 220 (46.2) | 8156 (49.7) |
| Supplement Use | 228 (47.9) | 7243 (44.1) |
| Current Smoker | 99 (20.8) | 3166 (19.3) |
| Alcohol | | |
|     <1 drink/week | 340 (71.4) | 11303 (68.9) |
|     1-7 drink/week | 85 (17.9) | 3255 (19.8) |
|     7+ drink/week | 51 (10.7) | 1787 (10.9) |
| Diabetes[1] | 40 (8.4) | 3218 (19.6) |
| Family history of Diabetes | 211 (44.3) | 7281 (44.4) |
| Hypertension[1] | 115 (24.2) | 4476 (27.3) |
| High cholesterol[1] | 189 (39.7) | 7394 (45.0) |

1. Background=Hispanic/Latino background; GED = General Education Development; Currently employed defined as part time or full time; Diabetes defined as either fasting time > 8 and fasting glucose ≥ 126 mg/dL, fasting time ≤ 8 and fasting glucose ≥ 200 mg/dL, glucose post- oral glucose tolerance test (OGTT) ≥ 200 mg/dL, % glycosylated hemoglobin (A1C) ≥ 6.5%, or use of anti-diabetic medication; Hypertension defined as systolic or diastolic BP ≥ 140/90 or use of antihypertensive mediations. High cholesterol defined as either total cholesterol ≥ 240 mg/dL, LDL-cholesterol ≥ 160 mg/dL, HDL-cholesterol ≥ 0 and < 40 mg/dL, or use of antihyperlipidemic medication.
2. Percentages may not add up to 100 because of missing data. In SOLNAS: currently employed (N=1), current smoker (N=1), family history of diabetes (N=2).
3. In HCHS/SOL: background (N=87), body mass index (N=71), education (N=91), currently employed (N=306), supplement use (N=630), current smoker (N=93), alcohol (N=70), diabetes (N=21), family history of diabetes (N=117), hypertension (N=2), high cholesterol (N=21).

**Table 2.** Age-Adjusted Geometric Mean Values for Nutritional Biomarker and Self-Reported Measures of α-carotene, α-tocopherol, vitamin B12, β-carotene, β-cryptoxanthin, Folate, γ-tocopherol, Lycopene, Retinol, Retinyl Palmitate, and Zeaxanthin by sex, shown for the mean age in SOLNAS, 46.1 years (N=476[1]).

|  | Female (N = 187) | | Male (N = 289) | |
|---|---|---|---|---|
|  | Mean | 95% CI | Mean | 95% CI |
| α-carotene | | | | |
|     2-Day mean, 10 mcg | 5.68 | 4.49, 7.19 | 5.90 | 4.40, 7.92 |
|     Biomarker, mcg/dL | 4.02 | 3.64, 4.44 | 2.75 | 2.43, 3.12 |
| α-tocopherol | | | | |
|     2-Day mean, mg | 6.56 | 6.12, 7.03 | 8.72 | 7.99, 9.51 |
|     Biomarker, mg/dL | 0.87 | 0.84, 0.90 | 0.87 | 0.83, 0.90 |
| Vitamin B12 | | | | |
|     2-Day mean, mcg | 2.92 | 2.68, 3.19 | 3.76 | 3.38, 4.19 |
|     Biomarker, $10^2$ pg/ml | 5.60 | 5.32, 5.89 | 5.12 | 4.81, 5.45 |
| β-carotene | | | | |
|     2-Day mean, $10^2$ mcg | 8.27 | 7.15, 9.57 | 8.34 | 6.96, 9.99 |
|     Biomarker, 10 mcg/dL | 1.29 | 1.18, 1.41 | 0.82 | 0.73, 0.92 |
| β-cryptoxanthin | | | | |
|     2-Day mean, 10 mcg | 2.81 | 2.34, 3.37 | 2.45 | 1.96, 3.08 |
|     Biomarker, mcg/dL | 8.79 | 8.01, 9.64 | 6.51 | 5.80, 7.31 |
| Folate | | | | |
|     2-Day mean, $10^2$ mcg | 3.13 | 2.95, 3.33 | 4.19 | 3.89, 4.51 |
|     Biomarker, 10 nmol/dL | 4.31 | 4.11, 4.51 | 4.18 | 3.94, 4.42 |
| γ-tocopherol | | | | |
|     2-Day mean, 10 mg | 0.71 | 0.65, 0.77 | 0.96 | 0.86, 1.07 |
|     Biomarker, mg/dL | 0.19 | 0.18, 0.19 | 0.17 | 0.16, 0.18 |
| Lycopene | | | | |
|     2-Day mean, $10^2$ mcg | 2.90 | 2.06, 4.08 | 5.43 | 3.54, 8.31 |
|     Biomarker, 10 mcg/dL | 3.08 | 2.91, 3.25 | 3.26 | 3.04, 3.49 |
| Retinol | | | | |
|     2-Day mean, $10^2$ mcg | 2.15 | 1.92, 2.41 | 2.71 | 2.35, 3.11 |
|     Biomarker, mcg/dL | 0.49 | 0.47, 0.51 | 0.56 | 0.53, 0.59 |
| Retinyl palmitate[2] | | | | |
|     Biomarker, mcg/dL | 0.11 | 0.08, 0.15 | 0.10 | 0.08, 0.14 |
| Zeaxanthin | | | | |
|     2-Day mean, $10^2$ mcg | 4.94 | 4.35, 5.61 | 5.34 | 4.56, 6.26 |
|     Biomarker, 10 mcg/dL | 1.71 | 1.61, 1.81 | 1.78 | 1.65, 1.91 |

1. N=476 for biomarker means with the exception of the missing data: vitamin B12 (N=5), Retinyl palmitate (N=221), Total Folate (N=11), Retinol (N=18); N=474 for self-reported measures due to two 24-hour recalls that listed an energy intake of <500 kcal/day and were deleted.
2. No exact 24-hour recall corresponding retinyl palmitate existed in data.

**Table 3.** Age-and-Sex-Adjusted Geometric Mean Values for Nutritional Biomarker and Self-Reported Measures of α-carotene, α-tocopherol, vitamin B12, β-carotene, β-cryptoxanthin, Folate, γ-tocopherol, Lycopene, Retinol, Retinyl Palmitate, and Zeaxanthin by Hispanic/Latino Background, shown for females and the mean age in SOLNAS, 46.1 years. (N=476[1]).

| | Central American (N = 51) | | Cuban (N = 69) | | Dominican (N = 48) | | Mexican (N = 142) | | Puerto Rican (N = 123) | | South American (N = 43) | |
|---|---|---|---|---|---|---|---|---|---|---|---|---|
| | Mean | 95% CI | Mean | 95% CI | Mean | 95% CI | Mean | 95% CI | Mean | 95% CI | Mean | 95% CI |
| α-carotene | | | | | | | | | | | | |
|   2-Day mean, $10^2$ mcg | 0.69 | 0.39, 1.23 | 0.69 | 0.41, 1.15 | 0.91 | 0.50, 1.63 | 0.59 | 0.42, 0.84 | 0.28 | 0.19, 0.42 | 1.13 | 0.61, 2.09 |
|   Biomarker, mcg/dL | 5.51 | 4.42, 6.87 | 4.18 | 3.43, 5.09 | 8.27 | 6.60, 10.36 | 3.96 | 3.46, 4.54 | 2.26 | 1.95, 2.63 | 5.83 | 4.60, 7.39 |
| α-tocopherol | | | | | | | | | | | | |
|   2-Day mean, mg | 6.09 | 5.15, 7.20 | 6.48 | 5.57, 7.52 | 4.45 | 3.74, 5.29 | 7.86 | 7.09, 8.72 | 6.18 | 5.52, 6.92 | 6.86 | 5.73, 8.22 |
|   Biomarker, mg/dL | 0.85 | 0.79, 0.92 | 0.83 | 0.77, 0.89 | 0.87 | 0.80, 0.94 | 0.94 | 0.90, 0.99 | 0.81 | 0.76, 0.85 | 0.85 | 0.78, 0.93 |
| Vitamin B12 | | | | | | | | | | | | |
|   2-Day mean mcg | 2.57 | 2.08, 3.18 | 2.66 | 2.21, 3.22 | 2.34 | 1.88, 2.91 | 3.39 | 2.98, 3.86 | 2.97 | 2.57, 3.42 | 2.77 | 2.20, 3.48 |
|   Biomarker, $10^2$ pg/ml | 5.59 | 4.95, 6.31 | 4.83 | 4.33, 5.38 | 5.09 | 4.49, 5.76 | 6.22 | 5.76, 6.71 | 5.44 | 5.01, 5.90 | 5.79 | 5.08, 6.60 |
| β-carotene | | | | | | | | | | | | |
|   2-Day mean, $10^3$ mcg | 0.87 | 0.61, 1.24 | 0.79 | 0.58, 1.09 | 0.85 | 0.59, 1.22 | 1.04 | 0.83, 1.29 | 0.55 | 0.43, 0.70 | 1.08 | 0.74, 1.59 |
|   Biomarker, 10 mcg/dL | 1.34 | 1.09, 1.66 | 1.17 | 0.97, 1.41 | 1.78 | 1.43, 2.21 | 1.60 | 1.41, 1.82 | 0.81 | 0.70, 0.93 | 1.60 | 1.27, 2.01 |
| β-cryptoxanthin | | | | | | | | | | | | |
|   2-Day mean, 10 mcg | 3.09 | 2.03, 4.72 | 1.52 | 1.04, 2.22 | 2.12 | 1.37, 3.29 | 5.51 | 4.24, 7.15 | 1.51 | 1.13, 2.01 | 3.83 | 2.43, 6.04 |
|   Biomarker, 10 mcg/dL | 0.85 | 0.70, 1.03 | 0.64 | 0.54, 0.76 | 0.93 | 0.76, 1.13 | 1.48 | 1.31, 1.66 | 0.47 | 0.41, 0.53 | 1.16 | 0.94, 1.42 |
| Folate | | | | | | | | | | | | |
|   2-Day mean, $10^2$ mcg | 3.10 | 2.68, 3.58 | 3.08 | 2.70, 3.50 | 2.51 | 2.16, 2.91 | 3.56 | 3.25, 3.89 | 2.86 | 2.59, 3.16 | 3.40 | 2.90, 3.97 |
|   Biomarker, 10 nmol/dL | 4.27 | 3.82, 4.78 | 3.59 | 3.24, 3.98 | 4.26 | 3.80, 4.78 | 4.60 | 4.29, 4.93 | 4.31 | 3.99, 4.65 | 4.43 | 3.92, 4.99 |
| γ-tocopherol | | | | | | | | | | | | |
|   2-Day mean mg | 6.92 | 5.58, 8.59 | 7.12 | 5.87, 8.63 | 4.49 | 3.59, 5.61 | 8.55 | 7.49, 9.76 | 6.86 | 5.93, 7.93 | 6.57 | 5.21, 8.28 |
|   Biomarker mg/dL | 0.18 | 0.16, 0.20 | 0.19 | 0.18, 0.21 | 0.16 | 0.15, 0.18 | 0.21 | 0.19, 0.22 | 0.17 | 0.16, 0.18 | 0.19 | 0.17, 0.21 |
| Lycopene | | | | | | | | | | | | |
|   2-Day mean, $10^2$ mcg | 2.08 | 0.89, 4.85 | 3.06 | 1.43, 6.51 | 1.42 | 0.59, 3.40 | 4.09 | 2.43, 6.89 | 2.71 | 1.53, 4.79 | 3.04 | 1.22, 7.54 |
|   Biomarker, 10 mcg/dL | 2.79 | 2.44, 3.20 | 3.25 | 2.88, 3.67 | 4.07 | 3.54, 4.67 | 2.95 | 2.71, 3.20 | 3.08 | 2.81, 3.38 | 2.69 | 2.33, 3.12 |
| Retinol | | | | | | | | | | | | |
|   2-Day mean, $10^2$ mcg | 1.79 | 1.29, 2.24 | 1.90 | 1.48, 2.43 | 1.73 | 1.30, 2.30 | 2.53 | 2.14, 3.00 | 2.28 | 1.90, 2.75 | 2.01 | 1.49, 2.70 |
|   Biomarker, mcg/dL | 0.49 | 0.44, 0.54 | 0.54 | 0.49, 0.59 | 0.47 | 0.42, 0.52 | 0.50 | 0.46, 0.53 | 0.46 | 0.43, 0.49 | 0.48 | 0.43, 0.54 |
| Retinyl palmitate[2] | | | | | | | | | | | | |
|   Biomarker, mcg/dL | 0.12 | 0.06, 0.22 | 0.17 | 0.10, 0.28 | 0.03 | 0.01, 0.06 | 0.22 | 0.15, 0.31 | 0.04 | 0.03, 0.06 | 0.10 | 0.06, 0.19 |
| Zeaxanthin | | | | | | | | | | | | |
|   2-Day mean, $10^2$ mcg | 5.45 | 4.05, 7.34 | 3.43 | 2.63, 4.48 | 2.99 | 2.20, 4.07 | 7.70 | 6.41, 9.25 | 3.55 | 2.91, 4.35 | 6.34 | 4.60, 8.74 |
|   Biomarker, 10 mcg/dL | 1.86 | 1.63, 2.14 | 1.52 | 1.34, 1.71 | 1.93 | 1.68, 2.22 | 2.00 | 1.84, 2.18 | 1.31 | 1.19, 1.43 | 1.84 | 1.59, 2.13 |

1. N=476 for biomarker means with the exception of the missing data: Vitamin B12 (N=5), Retinyl palmitate (N=221), Total Folate (N=11), Retinol (N=18); N=474 for self-reported measures due to two 24-hour recalls that listed an energy intake of <500 kcal/day and were deleted.
2. No exact 24-hour recall corresponding retinyl palmitate existed in data.

**Table 4.** Regression calibration β coefficients for logarithm of biomarker using self-reported intake and other important subject characteristics. Regressions are done on the log-transformed measures (N=450[1]).

| Variable | α-carotene (N = 450) | | α-tocopherol (N = 450) | | Vitamin B12 (N = 407) | | β-carotene (N = 401) | | β-cryptoxanthin (N = 450) | |
|---|---|---|---|---|---|---|---|---|---|---|
| | β | SE | β | SE | β | SE | β | SE | β | SE |
| Intercept[2] | 0.051 | 0.279 | -0.668 | 0.109 | 6.428 | 0.112 | 1.085 | 0.336 | 1.287 | 0.247 |
| Self-report | 0.096 | 0.017 | 0.02 | 0.022 | 0.032 | 0.031 | 0.105 | 0.030 | 0.112 | 0.020 |
| Background[3] | | | | | | | | | | |
|   Central Am.[4] | 0.334 | 0.124 | -0.095 | 0.043 | -0.115 | 0.082 | -0.18 | 0.129 | -0.501 | 0.108 |
|   Cuban | 0.174 | 0.115 | -0.109 | 0.042 | -0.226 | 0.077 | -0.124 | 0.117 | -0.574 | 0.104 |
|   Dominican | 0.780 | 0.137 | -0.066 | 0.046 | -0.213 | 0.085 | 0.146 | 0.14 | -0.339 | 0.12 |
|   Puerto Rican | -0.208 | 0.099 | -0.099 | 0.035 | -0.117 | 0.066 | -0.391 | 0.102 | -0.783 | 0.089 |
|   South Am.[5] | 0.322 | 0.13 | -0.124 | 0.046 | -0.071 | 0.084 | -0.06 | 0.132 | -0.247 | 0.113 |
| BMI[6] | -0.018 | 0.006 | 0.002 | 0.002 | -0.003 | 0.004 | -0.022 | 0.006 | -0.013 | 0.005 |
| Age[7] | 0.009 | 0.003 | 0.006 | 0.001 | -0.001 | 0.002 | 0.007 | 0.003 | 0.000 | 0.003 |
| Female | 0.332 | 0.085 | -0.012 | 0.03 | 0.04 | 0.053 | 0.41 | 0.086 | 0.142 | 0.074 |
| Supp. Use[8] | 0.087 | 0.069 | 0.045 | 0.025 | 0.104 | 0.046 | 0.149 | 0.071 | 0.085 | 0.060 |

| Variable | Folate (N = 408) | | γ-tocopherol (N = 450) | | Lycopene (N = 401) | | Retinol (N = 432) | | Zeaxanthin (N = 401) | |
|---|---|---|---|---|---|---|---|---|---|---|
| | β | SE | β | SE | β | SE | β | SE | β | SE |
| Intercept[2] | 3.049 | 0.240 | -2.115 | 0.120 | 2.472 | 0.200 | -0.906 | 0.180 | 2.087 | 0.229 |
| Self-report | 0.090 | 0.039 | 0.015 | 0.022 | 0.011 | 0.008 | -0.001 | 0.019 | 0.084 | 0.022 |
| Background[3] | | | | | | | | | | |
|   Central Am.[4] | -0.076 | 0.07 | -0.155 | 0.057 | -0.078 | 0.091 | -0.021 | 0.064 | -0.083 | 0.089 |
|   Cuban | -0.182 | 0.067 | -0.105 | 0.055 | 0.061 | 0.082 | 0.093 | 0.061 | -0.157 | 0.082 |
|   Dominican | 0.013 | 0.071 | -0.214 | 0.061 | 0.39 | 0.096 | -0.089 | 0.069 | -0.039 | 0.096 |
|   Puerto Rican | -0.013 | 0.057 | -0.177 | 0.047 | 0.099 | 0.071 | -0.061 | 0.053 | -0.24 | 0.071 |
|   South Am.[5] | -0.019 | 0.072 | -0.116 | 0.061 | -0.144 | 0.093 | -0.074 | 0.067 | -0.123 | 0.091 |
| BMI[6] | -0.01 | 0.003 | 0.008 | 0.003 | -0.004 | 0.004 | -0.002 | 0.003 | -0.009 | 0.004 |
| Age[7] | 0.004 | 0.002 | 0.001 | 0.002 | -0.003 | 0.002 | 0.001 | 0.002 | 0.001 | 0.002 |
| Female | 0.058 | 0.045 | 0.055 | 0.040 | 0.017 | 0.059 | -0.065 | 0.045 | 0.004 | 0.059 |
| Supp. Use[8] | 0.109 | 0.042 | -0.096 | 0.033 | -0.034 | 0.050 | 0.072 | 0.037 | -0.017 | 0.048 |

1. N=450 participants had complete baseline and cholesterol variable data; some prediction models included fewer participants due to the following missing variables in SOLNAS: current smoker (N=1), family history of diabetes (N=2), vitamin B12 biomarker (N=5), folate biomarker (N=11), retinol biomarker (N=18), LLD medication (N=10), HDL cholesterol (N=1), LDL cholesterol (N=12), vitamin B12 supplement intake (N=53), β-carotene supplement intake (N=49), folate supplement intake (N=49), lycopene supplement intake (N=49), zeaxanthin supplement intake (N=49).
2. The intercept represents the mean for the baseline group of nonsmoking, diabetes, and hypertension-free males of average age (46.1 years), average body mass index of 29.6, Spanish language preference, Mexican background, low income (< $20,000), low alcohol use (<1 drink/week), and less than high school education who do not have high cholesterol, have no family history of diabetes, and who do not take supplements.
3. Hispanic/Latino Background
4. Central American
5. South American
6. Body Mass Index (kg/m$^2$), centered at mean (29.6)
7. Age is centered at mean 46.1 years
8. Supplement Use

**Table 5**. Mean and median % bias, average standard errors (ASE), empirical standard errors (ESE), coverage probabilities (CP), and power for β-cryptoxanthin, Lycopene, and Folate in a time-to-event model when the true $\beta$ corresponding to the nutrient measure in the outcome model is assumed to be log(0.862) = -0.149 for β-cryptoxanthin, log(0.451) = -0.796 for Lycopene, and log(0.651) = -0.429 for Folate, which correspond to hazard ratios for a 20% increase in true nutrient consumption of 0.973, 0.865, and 0.925, respectively. Results are based on 2500 simulations and 1000 bootstrap replications.

| B-cryptoxanthin ($R^2 = 0.5035$) | Mean % Bias | Median % Bias | ASE[1] | ESE[2] | CP | Power |
|---|---|---|---|---|---|---|
| Truth[3] | 0.279 | 0.481 | 0.076 | 0.075 | 0.952 | 0.921 |
| Naïve Biomarker[4] | -12.755 | -9.238 | 0.426 | 0.432 | 0.951 | 0.079 |
| Calibrated Biomarker[5] | 1.818 | 5.886 | 0.524 | 0.507 | 0.948 | 0.082 |
| Naïve Self-Report[6] | -88.82 | -88.933 | 0.012 | 0.012 | 0.000 | 0.689 |
| Calibrated Self-Report[7] | 0.151 | -0.331 | 0.107 | 0.105 | 0.948 | 0.679 |
| Optimal[8] | 0.061 | -0.671 | 0.104 | 0.105 | 0.946 | 0.698 |
| **Lycopene** ($R^2 = 0.2196$) | Mean % Bias | Median % Bias | ASE | ESE | CP | Power |
| Truth[3] | 0.515 | 0.948 | 0.198 | 0.197 | 0.950 | 0.899 |
| Naïve Biomarker[4] | -33.613 | -29.26 | 0.988 | 1.013 | 0.945 | 0.074 |
| Calibrated Biomarker[5] | -1.448 | 2.623 | 1.583 | 1.522 | 0.947 | 0.074 |
| Naïve Self-Report[6] | -98.952 | -98.973 | 0.006 | 0.006 | 0.000 | 0.184 |
| Calibrated Self-Report[7] | 3.357 | -1.302 | 0.997 | 0.657 | 0.952 | 0.178 |
| Optimal[8] | -2.100 | -3.186 | 0.624 | 0.589 | 0.963 | 0.150 |
| **Folate** ($R^2 = 0.1717$) | Mean % Bias | Median % Bias | ASE | ESE | CP | Power |
| Truth[3] | 0.254 | 0.425 | 0.126 | 0.125 | 0.950 | 0.904 |
| Naïve Biomarker[4] | -47.812 | -45.279 | 0.56 | 0.581 | 0.929 | 0.069 |
| Calibrated Biomarker[5] | -1.288 | -2.872 | 1.253 | 1.119 | 0.944 | 0.072 |
| Naïve Self-Report[6] | -91.047 | -91.175 | 0.037 | 0.037 | 0.000 | 0.167 |
| Calibrated Self-Report[7] | 4.986 | -2.230 | 2.116 | 0.476 | 0.951 | 0.166 |
| Optimal[8] | -3.132 | -7.529 | 0.465 | 0.415 | 0.969 | 0.099 |

1. ASE is defined as the mean of the estimated standard errors from the model or bootstrap standard errors. For the calibrated approaches, ASEs are calculated as bootstrap standard errors while coverage probability and power are calculated from a bootstrap percentile confidence interval.
2. ESE is the empirical standard deviation of the estimated coefficients across simulations.
3. True biomarker average level used as exposure in outcome model fit to entire cohort (N=16,415); maximum power achieved.
4. Observed biomarker (subject to random, classical measurement error) used as exposure in outcome model fit to only the biomarker sub-study data (N=476).
5. Predicted biomarker level, adjusted for measurement error, used as exposure in outcome model fit to only the biomarker sub-study data (N=476).
6. Observed self-reported dietary variable (subject to systematic and random error) used as exposure in outcome model fit to entire cohort (N=16,415).
7. Predicted self-report, adjusted for measurement error, used as exposure in outcome model fit entire cohort (N=16,415).
8. An optimal combination of the calibrated biomarker and calibrated self-report; theoretically can have better power than Naïve and Calibrated Self-Report.

**Figure 1.** Comparison of the main study and reliability measures for each biomarker measurement (N=95). The dotted line denotes the 45-line (y=x). Each plot gives Pearson correlations for the logarithm of Visit 1 measures versus Visit 3 measures.

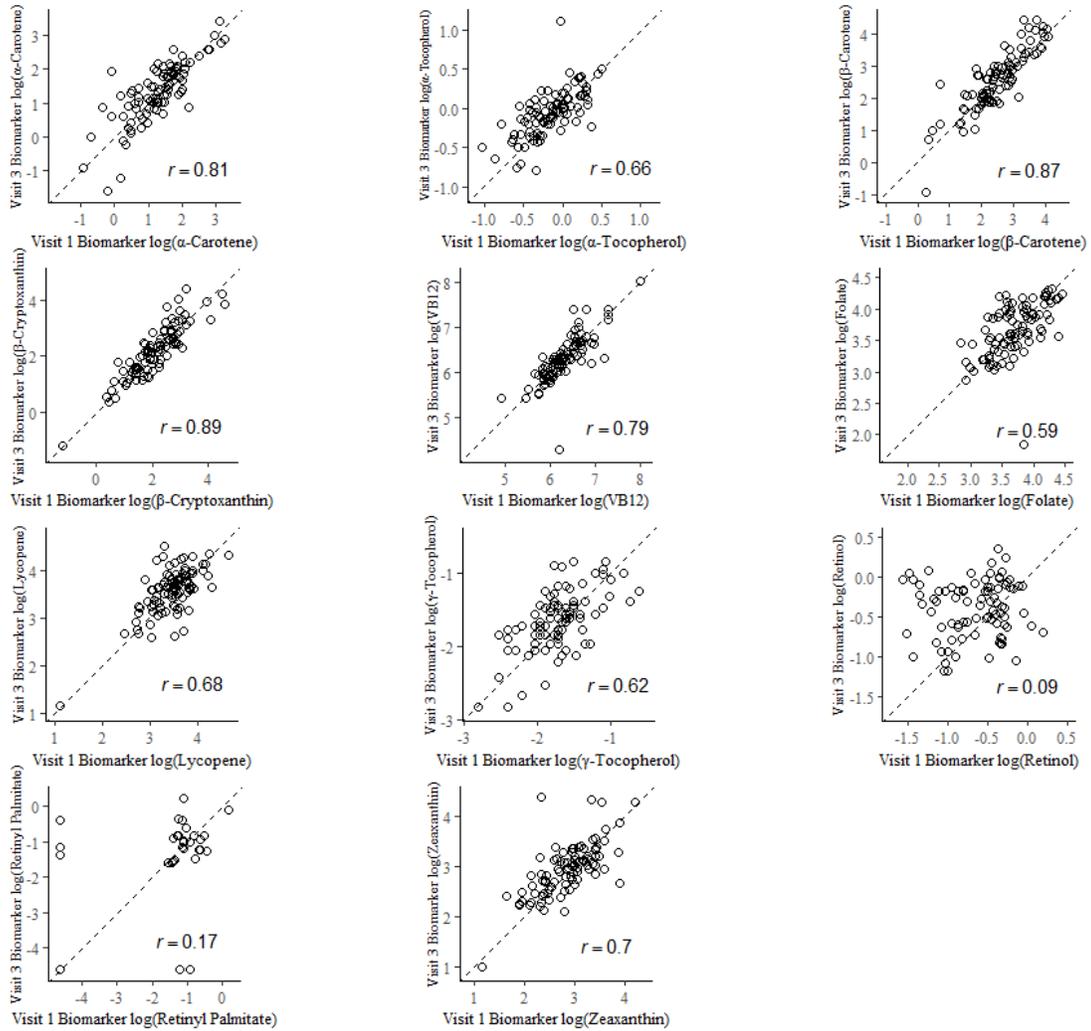

**Figure 2.** Comparison of the main study and reliability measures for each self-reported 24-hour Recall measurement (N=95). The dotted line denotes the 45-line (y=x). Each plot gives Pearson correlations for the logarithm of Visit 1 measures versus Visit 3 measures.

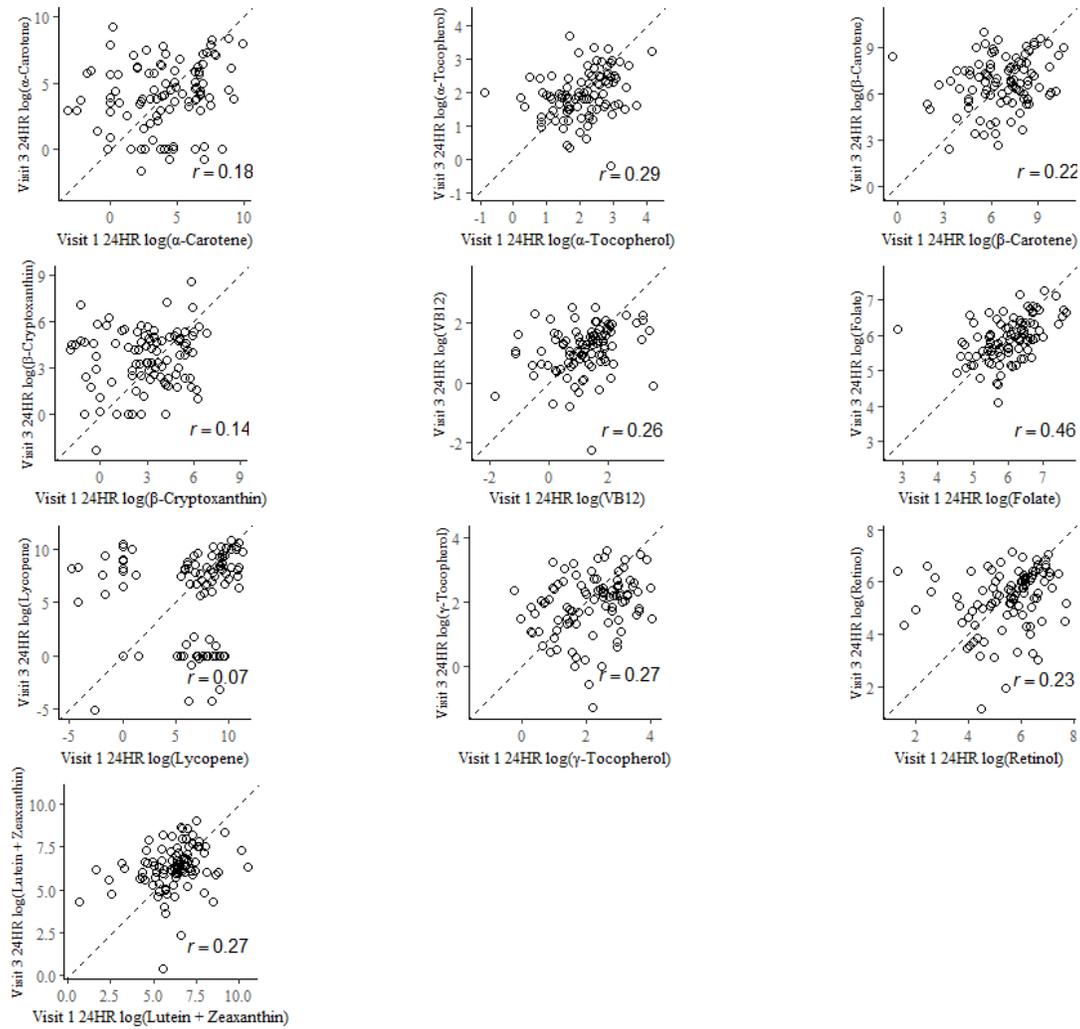

# Web Material

Blood concentration biomarkers for diet in in the Hispanic Community Health Study/Study of Latinos: Measurement characteristics and power

# Content

**Web Appendix 1: Supplemental details on study protocol and procedures**

*Serum Biomarkers*

As described in the paper, serum cholesterol measures were analyzed by the same central laboratory from a fasting blood draw taken at HCHS/SOL. Total cholesterol was measured by a Roche Modular P Chemistry Analyzer (Roche Diagnostics Corporation) using a cholesterol oxidase enzymatic method (Roche Diagnostics, Indianapolis, IN 46250). HDL-Cholesterol was measured in serum on a Roche Modular P Chemistry Analyzer (Roche Diagnostics Corporation) using a direct magnesium/dextran sulfate method (Roche Diagnostics, Indianapolis, IN 46250) and LDL was derived analytically using the Friedwald equation. Details of the lab analytes are available on the online under Manual 7 - Addendum Central Laboratory Procedures on the HCHS/SOL website.[1]

*Dietary and participant characteristics assessment*

24-hour dietary recalls in this study were conducted using the Nutrition Data System for Research software (version 11) developed by the Nutrition Coordinating Center, University of Minnesota, (Minneapolis, Minnesota), which uses a multi-pass system. Dietary recalls were led by bilingual interviewers using the language that each respondent preferred. Further details are provided by Siega-Riz et al. (2014)[2]. The telephone recalls from the HCHS/SOL baseline period and the SOLNAS in-person 24-hour dietary recall were used to create the self-reported 2-day mean in this analysis because these data were recorded at the closest point in time to the serum and plasma biomarkers (see Web Figure 1).

We now define a few key participant characteristics in the HCHS/SOL cohort. Participants were defined as supplement users (yes/no) if at least one supplement was reported in the past 30 days from the 30-day NDSR Dietary Supplement Assessment Module conducted at HCHS/SOL baseline clinic visit. Hypertension was determined to be present if a systolic/diastolic BP of greater than or equal to 140/90 was recorded or if the participant reported use of antihypertensive mediations. Participants were recorded as having high cholesterol if any of the following were true: total cholesterol $\geq$ 240 mg/dL, LDL-cholesterol $\geq$ 160 mg/dL, HDL-cholesterol $\geq$ 0 and < 40 mg/dL, or participant reported use of antihyperlipidemic medication. Diabetes was determined using the American Diabetes Association definition[3] such that the disease was present if the any of the following were true (1) fasting time in hours > 8 AND fasting glucose $\geq$ 126 mg/dL; (2) fasting time in hours $\leq$ 8 AND fasting glucose $\geq$ 200 mg/dL; (3) post-oral glucose tolerance test (OGTT) glucose $\geq$ 200 mg/dL; (4) A1C$\geq$ 6.5% or (5) if scanned antidiabetics medication was reported.

**Web Appendix 2: Supplemental details on Statistical Analyses**

*Details on variables of interest for prediction models*

All prediction equations included the set of baseline variables outlined in Web Table 2: age, sex, background, language preference, BMI, education, income, smoking, alcohol, diabetes, family history of diabetes, hypertension, high cholesterol, log total energy intake, a binary supplement use indicator, the associated self-reported nutrient, and the corresponding measure of supplement use for each nutrient.

The prediction models for α-carotene, β-carotene, β-cryptoxanthin, lycopene, and zeaxanthin were adjusted for season, as studies have shown that there may be seasonal differences in concentrations of these particular nutrients.[4] Models for α-carotene, α-tocopherol, vitamin B12, β-carotene, β-cryptoxanthin, and zeaxanthin were all adjusted for self-reported physical activity. In Lampe et al. (2016)[5], physical activity was only shown to only be an important predictor of vitamin B-12, but we chose to include it in the prediction models for a few other carotenoids and tocopherols that have been observed to be related to physical activity.[6,7] Previous literature shows that carotenoids and tocopherols need to be adjusted for cholesterol because blood lipids may interfere with concentration marker levels and confound their interpretation in disease models.[8] Rather than using existing equations to adjust these biomarker nutrient values, we include all cholesterol-related variables in our prediction models for carotenoids and tocopherols. Cholesterol variables consist of HDL-cholesterol (mg/dL), LDL-cholesterol (mg/dL), and lipid lowering drugs/ Antihyperlipidemics (LLD).

Goodman et al. (1996)[9] found associations between plasma α-tocopherol and self-reported dietary β-carotene, plasma β-carotene and self-reported dietary α-tocopherol, and plasma retinol and self-reported dietary β-carotene. Thus, prediction models for α-tocopherol, β-carotene, and retinol were further adjusted to include other self-reported nutrient values beyond the exact corresponding measure (Web Table 3). Specifically, models for α-tocopherol and retinol were also adjusted for self-reported β-carotene. Additionally, the prediction model for β-carotene was also adjusted for self-reported α-tocopherol.

**Web Appendix 3: Procedures to check for overfitting**

For each prediction model, we check for overfitting using a procedure proposed by Harrell (2015)[10]. For this assessment, we will consider $R^2$ values as an accuracy index but will use the bootstrap to approximate out-of-sample predictive accuracy. In this procedure, we draw a sample of our observations with replacement of the same size as our data used to fit our prediction model. Then, we fit a model using this bootstrap sample and apply it to the original data. This procedure enables us to estimate the bias resulting from overfitting, which is referred to as the "optimism" of the fitted model.[10] The optimism estimate is calculated as the $R^2$ from the bootstrap model minus the $R^2$ from the original sample. We repeat this process for 100 bootstrap replications and obtain an average optimism, which we subtract from the $R^2$ of the final model in order to obtain an index corrected for overfitting. This process can be implemented easily using the `rms` package in R software.[11] If overfitting is not a concern, the $R^2$ corrected for overfitting should closely resemble the original $R^2$ from our prediction model. To evaluate overfitting, we consider anything less than an absolute change in the $R^2$ of 0.1 to be acceptable, 0.1-0.2 to be a cause for some concern, and greater than 0.2 to be extremely concerning. If there is evidence of overfitting, we consider two methods for variable reduction.[10]

To fit reduced models in the presence of overfitting, we first consider a redundancy analysis procedure for constructing reduced calibration models for each nutrient based on how well a variable can be predicted from other variables in the model. The procedures are implemented using the *redun* function in the `Hmisc` package in R.[11] This process expands continuous predictors into restricted cubic spline basis functions and expands categorical predictors into dummy variables. Ordinary least squares is then used to predict each individual predictor with all components of the remaining predictors. Any variable that can be predicted with a high $R^2$ value by the remaining set of predictors is removed. We continue this process for all other predictors until no single predictor can be predicted with an $R^2$ of at least *r* or until dropping a predictor causes a variable that was dropped earlier to no longer be able to be predicted at level *r* from the reduced list of variables. We choose *r* = 0.80 for our threshold.

A second procedure for data reduction is variable clustering, which can be used to determine the relationship between candidate predictor variables. Variable clustering is an approach to assessing collinearity and redundancy as well as creating clusters of variables that may be treated as a single variable. The `Hmisc` *varclus* function in R performs a hierarchical cluster analysis on variables using a similarity matrix. For a similarity measure, we select the robust Hoeffding's D statistic which has the ability to identify dependencies between any two variables. Hoeffing's D provides a measure of the agreement between F(x,y) and G(x)H(y), where F is the joint cumulative distribution function (CDF) between two variables X and Y and G and H are marginal CDFs.

**Web Appendix 4: Details of numerical simulation**

Simulations were based on the HCHS/SOL cohort. We chose a population of similar size, with N = 16,415 for the full cohort, N = 476 for the SOLNAS sub-study, and N = 95 in the reliability study for which we have two biomarker measures available. We consider three separate outcome models for three nutrients: β-Cryptoxanthin, lycopene, and folate. For each model, we further assume that age and body mass index (BMI) are related to the outcome of interest. We generate three multivariate normal distributions, one for each nutrient, with means and covariance matrices based on the full HCHS/SOL cohort. For example, we used a mean of the log of the 2-day average of 24-hour recall measures of β-Cryptoxanthin, lycopene, and folate of 3.261392, 5.605585, and 5.736064, respectively, and a mean age of 45.81989 and mean BMI of 29.77589 to generate this data. Below are the covariance matrices used to generate our data, with the variance for the corresponding nutrient in position [1,1], the variance for age in position [2,2], and the variance for BMI in position [3,3]:

$$\Sigma_{\beta-\text{crypt}} = \begin{bmatrix} 2.7095730 & 0.5280317 & -0.4143209 \\ 0.5280317 & 194.0924000 & 8.3544090 \\ -0.4143209 & 8.3544090 & 36.8888900 \end{bmatrix},$$

$$\Sigma_{\text{Lycopene}} = \begin{bmatrix} 9.0749150 & -2.981719 & -0.6255943 \\ -2.9817190 & 194.092400 & 8.3544090 \\ -0.6255943 & 8.3544090 & 36.8888900 \end{bmatrix},$$

and

$$\Sigma_{\text{Folate}} = \begin{bmatrix} 0.2948574 & -0.6786461 & -0.2807268 \\ -0.6786461 & 194.092400 & 8.3544090 \\ -0.2807268 & 8.3544090 & 36.8888900 \end{bmatrix}.$$

To simulate the true nutrient measure from the simulated error-prone measure, we assumed the following model: $X = \alpha_0 + \alpha_1 X^* + \alpha_2 \text{AGE} + \alpha_3 \text{BMI} + u$, where X is the true nutrient measure, $X^*$ is the error-prone measure generated from the multivariate normal distribution described above, and u is mean zero, random error with variance $\sigma_u^2$. The $\alpha_0, \alpha_1, \alpha_2$, and $\alpha_3$ values used to simulate each true nutrient measure were based on the calibration coefficients presened in Table 4 of the main manuscript. To simulate the biomarker measure of each nutrient, we assumed the classical measurement error model as follows: $X^{**} = X + \epsilon$, where $X^{**}$ is the biomarker measure, X is the true nutrient measure simulated above, and $\epsilon$ is mean zero, random error with variance $\sigma_\epsilon^2$. We simulate the $R^2$ values based on the full prediction models in the data to be 0.5034792 for β Cryptoxanthin, 0.2196337 for Lycopene, and 0.1716752 for Folate. This is accomplished by simulating a total variance $\sigma_T^2 = \sigma_\epsilon^2 + \sigma_u^2$ as a function of the $R^2$. Note that the $R^2$ for each prediction model has the following form:

$$R^2 = \frac{\text{Var}(X^{**}) - \sigma_T^2}{\text{Var}(X^{**})}$$

We can rearrange this formula to solve for $\sigma_T^2$ as follows: $\sigma_T^2 = (1 - R^2)\text{Var}(X^{**})$. Lastly, to simulate $\sigma_T^2$, we must derive the variance of the biomarker, $X^{**}$, as below:

$$\begin{aligned}\text{Var}(X^{**}) &= \text{Var}(\alpha_0 + \alpha_1 X^* + \alpha_2 \text{AGE} + \alpha_3 \text{BMI} + u + \epsilon) \\ &= \alpha_1 \text{Var}(X^*) + \alpha_2 \text{Var}(\text{AGE}) + \alpha_3 \text{Var}(\text{AGE}) + \sigma_T^2 + 2\alpha_1\alpha_2 \text{Cov}(X^*, \text{AGE}) \\ &\quad + 2\alpha_1\alpha_3 \text{Cov}(X^*, \text{BMI}) + 2\alpha_2\alpha_3 \text{Cov}(\text{AGE}, \text{BMI}) = A + \sigma_T^2 \end{aligned}$$

where $A = \alpha_1 \text{Var}(X^*) + \alpha_2 \text{Var}(\text{AGE}) + \alpha_3 \text{Var}(\text{AGE}) + 2\alpha_1\alpha_2 \text{Cov}(X^*, \text{AGE}) + 2\alpha_1\alpha_3 \text{Cov}(X^*, \text{BMI}) + 2\alpha_2\alpha_3 \text{Cov}(\text{AGE}, \text{BMI})$

Now, we see that $\sigma_T^2 = (1 - R^2)\text{Var}(X^{**}) = (1 - R^2)(A + \sigma_T^2)$. Rearranging, we have the following:

$$\sigma_T^2 = \frac{A(1 - R^2)}{R^2}$$

which is used to simulate the $R^2$ values for the three regression calibration equations based on those observed in the data. We choose $\sigma_\epsilon^2$ based on the variance of repeat biomarker measures observed in the reliability subset. In particular, we found the ratio between the variance of repeat biomarker measures ($\sigma_{\epsilon,\text{Data}}^2$) and the prediction model residual variance from the data ($\sigma_{T,\text{Data}}^2$), and solved the equation below for $\sigma_\epsilon^2$, where the value $\sigma_{T,\text{SIM1}}^2$ was chosen based on the simulated value of $\sigma_T^2$ described above, but calculated from a single simulation iteration:

$$\frac{\sigma_{\epsilon,\text{Data}}^2}{\sigma_{T,\text{Data}}^2} = \frac{\sigma_\epsilon^2}{\sigma_{T,\text{SIM1}}^2}$$

Solving this equation resulted in a $\sigma_\epsilon^2$ of 0.01070956 for β-Cryptoxanthin, 0.004405707 for lycopene, and 0.02127483 for folate. Finally, at each iteration we simulate $\sigma_T^2$ and then calculate $\sigma_U^2 = \sigma_T^2 - \sigma_\epsilon^2$.

Lastly, we simulate the time-to-event from an exponential distribution with parameter $\lambda = \lambda_0 \exp(\beta_1 X + \beta_2 \text{AGE} + \beta_3 \text{BMI})$. The true regression parameters, $\beta_j, j = 1, 2, 3$, were based on a hypothesized strength of association between each exposure and diabetes that resulted in approximately 90% power $\beta_j$, . In particular, we chose $\beta_1$ to be $\log(0.775) = -0.255$ for β-cryptoxanthin, $\log(0.529) = -0.637$ for Lycopene, and $\log(0.665) = -0.408$ for Folate, which correspond to hazard ratios for a 20% increase in true nutrient consumption of 0.954, 0.890, and 0.928, respectively. Further, we selected $\beta_2 = \log(0.9)$ and $\beta_3 = \log(0.75)$. To mimic diabetes incidence in the HCHS/SOL cohort, we simulate a censoring rate of approximately 85% by selecting a final censoring time of 60 for all simulations, which is meant to represent time in months, and fixing $\lambda_0$ at 650, 2000, and 1600 for β-Cryptoxanthin, lycopene, and folate, respectively.

Finally, we note that for simplicitiy of presentation, we assumed HCHS/SOL was a simple random sample of US Hispanic/Latinos rather than a design-based sample.

**Web Appendix 5: Coefficient of variation calculation**

Coefficient of variation (CV) measures between the blind duplicate samples are estimated using linear mixed models with random intercepts, similarly to the approach described in Thyagarajan et al. (2016).[12] Let $Y_{ij}$ be defined as the jth biomarker measure for the ith participant. Then, we will assume the following random intercepts model:

$$Y_{ij} = \gamma_0 + b_{0i} + e_{ij},$$

where $\gamma_0$ is the intercept, $b_{0i}$ is the random effect term, and $e_{ij}$ is the random error term, which is assumed to be normally distributed with mean 0 and variance $\sigma_e^2$. The within-individual biomarker variance is estimated as the variance of $e_{ij}$. Then, for each biomarker, we compute the estimated coefficient of variation as the within-individual standard deviation (i.e. the square root of the within-individual biomarker variance) divided by the average value (i.e. the mean of the averages of the original and duplicate measures) and multiplied by 100.

**Web Table 1.** Intraclass correlation coefficients (ICC) and coefficient of variation (CV) measures between blind duplicate samples for blood biomarkers based on 10% of the cohort (n=49[1]).

| Nutrient | ICC[2] | CV[3] |
|---|---|---|
| α-carotene | 0.9826 | 12.7956 |
| α-tocopherol | 0.9119 | 9.1964 |
| Vitamin B12 | 0.9941 | 3.8048 |
| β-carotene | 0.9868 | 9.3896 |
| β-cryptoxanthin | 0.9585 | 13.9677 |
| Folate | 0.9779 | 4.9664 |
| γ-tocopherol | 0.8909 | 10.8904 |
| Lycopene | 0.9507 | 10.3379 |
| Retinol | 0.7660 | 16.4321 |
| Retinyl Palmitate | 0.6910 | 38.7439 |
| Zeaxanthin | 0.9131 | 13.9785 |

1. N=49 participants had blind duplicates for at least one biomarker with the exception of Vitamin B12 (N=48) and retinyl palmitate (N=29)
2. ICC measures are computed as the Pearson correlations between the blind duplicate samples for each biomarker.
3. For each sample, % CV measures are computed as the standard deviation (i.e. the square root of the within-individual variance component) divided by the average value (i.e. the mean of the averages of the original and duplicate measures), and multiplied by 100. (See Web Appendix 5 for details)

**Web Table 2**. Variables of interest in building prediction models for each nutrient biomarker.

| | α-carotene | α-tocopherol | Vitamin B12 | β-carotene | β-cryptoxanthin | Folate | γ-tocopherol | Lycopene | Retinol | Zeaxanthin |
|---|---|---|---|---|---|---|---|---|---|---|
| Baseline Variables[1] | × | × | × | × | × | × | × | × | × | × |
| Cholesterol Variables[2] | × | × | | × | × | | × | × | × | × |
| Season | × | | | × | × | | | × | | × |
| Physical Activity | × | × | × | × | × | | | | | × |

1. Baseline variables included in the prediction equations for all nutrients consist of age, sex, background, language preference, BMI, education, income, smoking, alcohol, diabetes, family history of diabetes, hypertension, high cholesterol, log total energy intake, a binary supplement use indicator, the associated self-reported nutrient, and the corresponding measure of supplement use for each nutrient.
2. Cholesterol variables consist of HDL-cholesterol (mg/dL), LDL-cholesterol (mg/dL), and lipid lowering drugs/Antihyperlipidemics (LLD).

**Web Table 3.** Nutrient biomarkers and corresponding self-reported nutrients for use in prediction models.

| Biomarker | Associated Self-Reported Nutrients |
|---|---|
| α-carotene (μg/dL) | α-carotene (provitamin A carotenoid) (mcg) |
| α-tocopherol (mg/dL) | Total α-tocopherol Equivalents (mg), β-carotene (provitamin A carotenoid) (mcg)[1] |
| Vitamin B12 (pg/ml) | Vitamin B12 (cobalamin) (mcg) |
| β-carotene (μg/dL) | β-carotene (provitamin A carotenoid) (mcg), Total α-tocopherol Equivalents (mg)[1] |
| β-cryptoxanthin (μg/dL) | β-cryptoxanthin (provitamin A carotenoid) (mcg) |
| Folate (nmol/L) | Total Folate (mcg) |
| γ-tocopherol (mg/dL) | γ-tocopherol (mg) |
| Lycopene (μg/dL) | Lycopene (mcg) (carotene) |
| Retinol (μg/dL) | Retinol (mcg), β-carotene (provitamin A carotenoid) (mcg)[1] |
| Zeaxanthin (μg/dL) | Lutein + Zeaxanthin (mcg) (xanthophyll or oxygenated carotenoids) |

1. Previous publications show associations between plasma α-tocopherol and self-reported dietary β-carotene; plasma β-carotene and self-reported dietary α-tocopherol; plasma retinol and self-reported dietary β-carotene.

**Web Table 4.** Variables chosen for the reduced prediction model for each nutrient by AIC in a stepwise selection algorithm, along with $R^2$ values for the full and reduced models.

| | α-carotene | α-tocopherol | Vitamin B12 | β-carotene | β-cryptoxanthin | Folate | γ-tocopherol | Lycopene | Retinol | Zeaxanthin |
|---|---|---|---|---|---|---|---|---|---|---|
| Corresponding Self-Report | × | | | × | × | × | | × | | × |
| Additional Self-Report | | × | | | | | | | × | |
| Age | × | × | | × | | × | | × | × | |
| Sex | × | | | × | × | | × | | × | |
| Background | × | × | × | × | × | × | × | × | × | × |
| Language Preference | × | | | × | × | | | | | × |
| BMI | × | | | × | × | × | × | × | × | × |
| Income | | | | × | | | | | | |
| Education | | | | | | | | | × | |
| Smoking | × | | | × | × | × | | | | |
| Alcohol | | | | | | | × | | × | |
| Diabetes | × | | | × | | | | | | |
| Family History of Diabetes | | | | | | | | | | |
| High Cholesterol | | | | | | | | | | |
| Hypertension | × | | | × | × | | | × | | |
| Log Total Energy Intake | | | × | | | | | × | | × |
| Supplement Use | × | × | × | × | × | × | × | | × | |
| Supplement Intake | | | × | | | × | | | | |
| HDL-Cholesterol | | | | | | | | | | |
| LDL-Cholesterol | × | × | | × | × | | × | × | | × |
| Lipid Lowering Drugs | | × | | | × | | | | × | × |
| Season | | | | × | | | | | | × |
| Physical Activity | | | | | × | | | | | |
| **$R^2$ Full** | 0.4692 | 0.3486 | 0.1013 | 0.4562 | 0.5035 | 0.1717 | 0.2340 | 0.2196 | 0.1185 | 0.2728 |
| **$R^2$ Stepwise** | 0.4518 | 0.3334 | 0.0797 | 0.4469 | 0.4856 | 0.1416 | 0.2211 | 0.1794 | 0.1060 | 0.2410 |

**Web Table 5.** $R^2$ values based on the full prediction models, Prentice $R^2$ values, Partial $R^2$ values, correlations between the biomarker and associated self-reported (SR) 2-day mean, and new $R^2$ measures that could exist if there were 2 and 4 repeat measures of each biomarker available on everyone in the SOLNAS sub-study.

| Nutrient | $R^2$ | Prentice $R^2$ | SR Partial $R^2$ | Correlation SR and Biomarker ($\rho$) | $R^2_{new}$ (2)[1] | $R^2_{new}$ (4)[1] |
|---|---|---|---|---|---|---|
| β-cryptoxanthin | 0.5035 | 0.5662 | 0.0699 | 0.4090 | 0.5329 | 0.5491 |
| α-carotene | 0.4692 | 0.5792 | 0.0725 | 0.3321 | 0.5183 | 0.5470 |
| β-carotene | 0.4562 | 0.5258 | 0.0326 | 0.2801 | 0.4872 | 0.5058 |
| α-tocopherol | 0.3486 | 0.5304 | 0.0019 | 0.0442 | 0.4205 | 0.4691 |
| Zeaxanthin | 0.2728 | 0.3918 | 0.0373 | 0.2447 | 0.3215 | 0.3532 |
| γ-tocopherol | 0.2340 | 0.3776 | 0.0011 | 0.0216 | 0.2889 | 0.3273 |
| Lycopene | 0.2196 | 0.3214 | 0.0046 | 0.1002 | 0.2607 | 0.2879 |
| Folate | 0.1717 | 0.2916 | 0.0137 | 0.1211 | 0.2154 | 0.2478 |
| Retinol | 0.1185 | 1.2871 | 0.0000 | 0.0236 | 0.2261 | 0.3847 |
| Vitamin B12 | 0.1013 | 0.1283 | 0.0028 | 0.0474 | 0.5606 | 0.5968 |

1. $R^2_{new}$ (j) is calculated as $R^2 \times \frac{Var(X^{**})}{Var(X^{**}_{new})} = R^2 \times \frac{Var(X^{**})}{Var(X)+Var(U)/j}$ and is interpreted as the new $R^2$ measure that would be available if there were $j$ repeat biomarker measures available on everyone in the SOLNAS sub-study. $Var(X)$ is the variance of the true nutrient measure calculated as $Var(X^{**}) \times ICC$, where $Var(X^{**})$ is the variance of the biomarker measure. $Var(U)$ is the variance of the measurement error of the biomarker.

**Web Figure 1.** Study timing and procedures in the SOL Nutrition & Physical Activity Assessment Study (SOLNAS) for blood concentration biomarkers (2010-2012).*

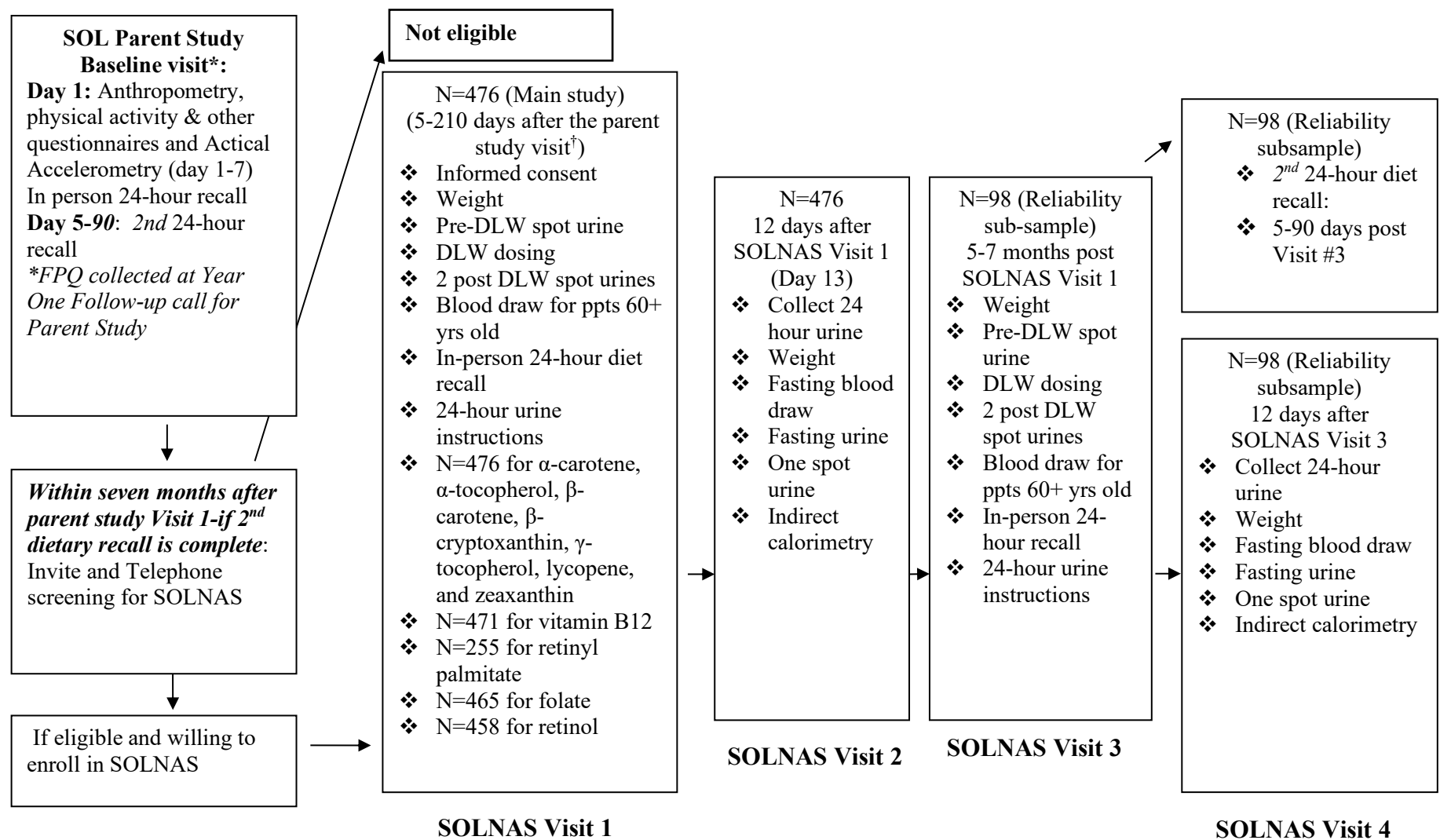

* Adapted from Mossavar-Rahmani et al. (2015) [13].
†For San Diego site only, SOLNAS Visit 1 window could be up to 12 months after parent study visit.

Hispanic Community Health Study/Study of Latinos. American Journal of Epidemiology. 2015 Jun 15;181(12):996-1007.